\documentclass[onecolumn]{els-mrw} 

\usepackage{amsmath,amssymb,amsfonts,amsthm,makeidx,graphicx}
\usepackage{txfonts}
\usepackage{helvet}

\begin{document}

\chapter{Quantum Chaos in Phase Space}\label{chap1}

\author[1]{Martina Hentschel}%


\address[1]{\orgname{Technische Universität Chemnitz}, \orgdiv{Institute for Physics}, \orgaddress{D-09107 Chemnitz, Germany}}

\articletag{Chapter Article tagline: Sept. 1, 2025}

\maketitle

\begin{glossary}[Keywords]
Quantum chaos, mesoscopic optics, Poincaré surface of section, ray-wave correspondence, trajectory tracing, phase space representation, Husimi functions, open billiards for light, anisotropic billiards, directional emission
\end{glossary}

\begin{abstract}[Abstract]
	Mesoscopic devices, with system sizes in the range of several to several dozens wavelengths, represent paradigmatic model systems for the observation of quantum chaotic behaviour based on semiclassical concepts.
	Those electronic and photonic billiard cavities are small enough for interference effects not to be ignored. Nonetheless, the classical ray or particle tracing picture can often provide a substantial understanding of the system’s dynamical behaviour along the lines of classical-quantum, or ray-wave correspondence. This well-established principle turns out to be particularly useful when applied not only in real space, but by extending it to phase space such that both location and momentum information can contribute to a deeper and more comprehensive understanding of the system’s dynamical behaviour, and explain the often observed universal nature of quantum chaotic systems. Noting that the role of Planck’s constant $\hbar$, the size of a phase space cell  in the description of electronic systems, is taken by the wavelength or inverse wavenumber in photonic systems highlights the conceptual merit of a phase-space description of the various mesoscopic systems.
\end{abstract}

\section{Introduction: From Chaos to Quantum Chaos} 
\label{chap1:sec1}

\subsection{The classical case: Nonlinear dynamics and chaos}
\begin{figure}[t]
\centering
\includegraphics[width=1.0\textwidth]{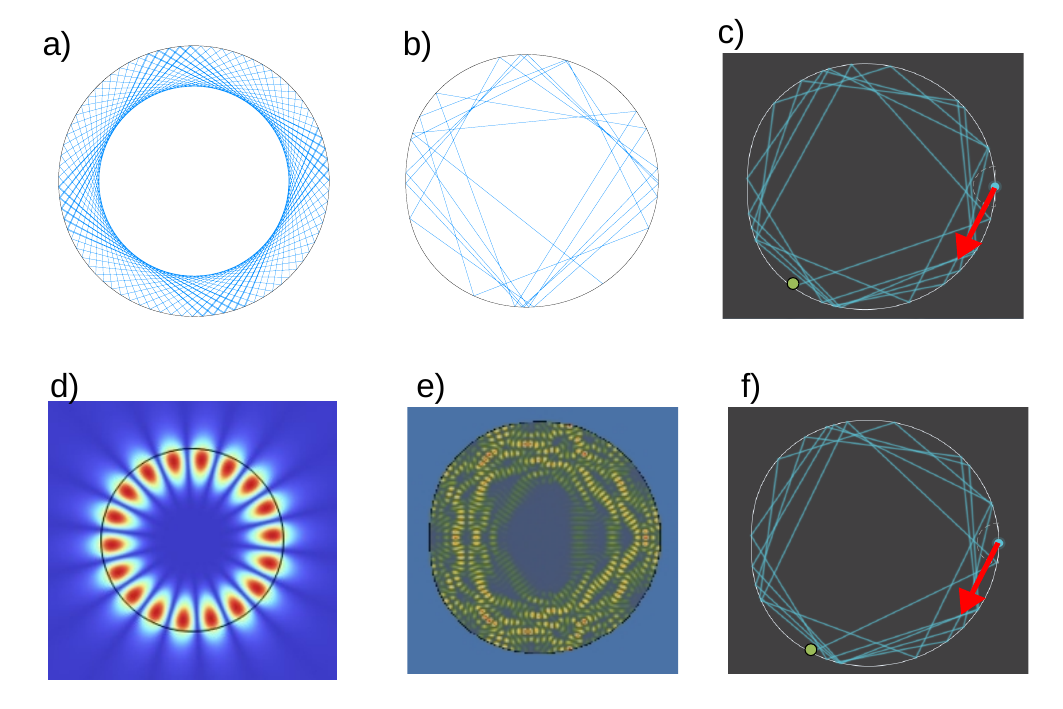} 
\caption{Examples for regular and chaotic dynamics. a) Regular trajectory of whispering-gallery type in a circular billiard, with d) a sample wave solution. b) Chaotic trajectory in a deformed cavity where angular momentum is not conserved, and e) a sample resonance solution. 
	c) and f) show chaotic particle/ray trajectories with very similar, yet nonidentical initial conditions (red arrows) that start to deviate after several reflections as indicated by the green dots. This illustrates the sensitive dependence on initial conditions for chaotic cavities.}
\label{fig_chaotictraj}
\end{figure}

When being asked about what chaos is all about, many people seem to think about the butterfly effect \cite{butterflyeffect}. 
It can be an impressive starting point, providing at the same time room for creative interpretations (with the observation of real butterflies justified in and of itself!). A more physical approach in the spirit of nonlinear dynamics is to characterize chaos as the exponential sensitivity of the system dynamics on the initial conditions: even tiny differences in the starting position and momentum vector can alter the dynamics not only quantitatively, but qualitatively after a short time. This is illustrated in Fig.~\ref{fig_chaotictraj} for the well-known example of billiard model systems. The trajectory in billiards \cite{Bunimovich}lows a straight line with specular reflection at the boundary. In the case of a curved boundary that is typically studied in physics, the tangent at the reflection point is considered. In Fig.~\ref{fig_chaotictraj}(a), a regular trajectory in a circular billiards is shown. The deformation of the circular cavity in Fig.~\ref{fig_chaotictraj}b) induces immediately chaotic trajectories. Their sensitive dependence on the initial conditions is illustrated in Fig.~\ref{fig_chaotictraj} c),d).   The divergence of two sample trajectories can be measured by the Lyapunov exponent \cite{lichtenberglieberman1992} in classical nonlinear dynamics. The study of trajectories in hard-wall billiard systems or in maps like the well-known baker or cat map, and following the (trajectory) dynamics in real space will provide a first impression of the system's dynamical character that can be regular or chaotic.


\subsection{The quantum case: signatures of classical chaos in the quantum world and quantum-classical correspondence}

Ray-wave or classical-quantum correspondence is at the foundation of quantum mechanics as it ensures the correct and smooth transition from the quantum world to the classical case for high energies/small wavelengths and large masses.
According to the paradigmatic principle of quantum-classical correspondence, we expect some relation between the classical and quantum behaviour. But there is no trajectory in quantum mechanics. Instead, the wave function describes the particle, and details below (half of) the wavelength cannot be resolved,
cf.~Fig.~\ref{fig_chaotictraj} d),e) that illustrate wave counterparts to the trajectories in Fig.~\ref{fig_chaotictraj} a),b). 
The sensitive dependence on initial conditions seems not to be a useful concept anymore (but we refer the reader to another chapter in this volume:  
Quantum analogues of exponential sensitivity: from Loschmidt echo to
Krylov complexity by Ignacio Garcia-Mata and Diego A. Wisniacki). 
For the study of the relativistic quantum case, we recommend the chapter 
Relativistic Quantum Chaos in Neutrino Billiards by Barbara Dietz in this volume where it 
is discussed in detail. 
The paradigmatic case of the kicked rotor is subject of the chapter 
The Quantum Kicked Rotor: A Paradigm of Quantum Chaos. Foundational aspects and new perspectives written by
Giuliano Benenti, Giulio Casati, Jiangbin Gong and Zhixing Zoud. 

Nonetheless, there are differences in the quantum mechanics of systems with regular and chaotic classical analogue, respectively  \cite{Stoeckibuch,Haakebuch,Takabuch,celso,raizen,gutzwiller_scholarpedia,ChaosBook}. 
An important role play the periodic orbits of the system, cf.~the chapter The role of classical periodic orbits in quantum many-body systems by 
Daniel Waltner and Boris Gutkin. In short, the entity of all periodic orbits of the classical pendant has to be considered and describes, via the Berry-Tabor \cite{berrytabortrace} and the Gutzwiller \cite{gutzwillertrace} trace formula, for integrable and chaotic systems respectively, the entity of all quantum states of the system in form of the density of states. We point out that there are many other quantities that capture the differences in more detail and that are the topic of this volume. 
A \emph{qualitative} similarity between single trajectories and individual resonances, along the lines of classical-quantum or ray-wave correspondence, can be found and is illustrated in a number of  examples. In this context the semiclassical limit of very small wavelength is of particular interest and briefly discussed below. 

\subsection{The hybrid case: classical trajectories carrying the essence of the quantum system's information }

We mention a third situation here: The great advantage of classical particle tracing modeling is the easy and (practically) barrier-free numerical access. These advantages are maintained when the quantum properties of a system can effectively be taken into account. One example are open quantum systems, like quantum dot billiard systems where the particles might escape just as in billiards for light. While this escape is captured by the well-known Fresnel transmission coefficients in the photonic case, generalized Fresnel coefficients have to be obtained for the generic situation, for example from wave matching at the boundary. Taking the quantum mechanical properties of the system properly into account in this way, allows for a particle-tracing description of the system that captures its qualitative behaviour. One example are bilayer graphene billiards \cite{BLG1,BLG2}, and we consider these as an example in more detail in the next section.  


The outline of this chapter is as follows. We next introduce Husimi functions that are an useful tool in mapping wave functions from real space into phase space. We discuss a number of phenomena where the phase-space point of view is particularly useful, and end with a brief summary. 

\section{Extending the consideration into phase space: Poincaré surface of section}
\begin{figure}[t]
	\includegraphics[width=1.0\textwidth]{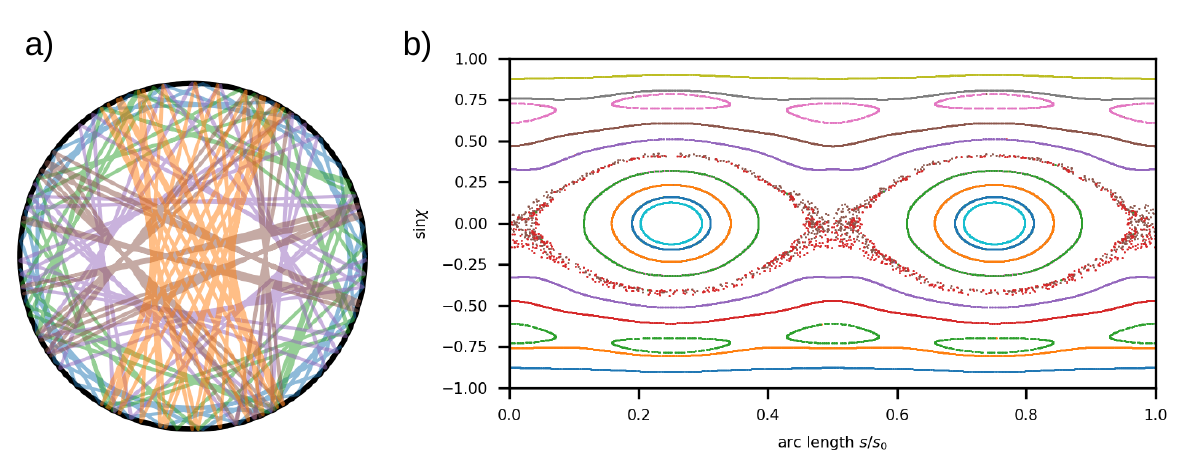}
	\label{fig_phsp_class}
\caption{a) Sample trajectories in real space for quadrupolar geometry, $r(\varphi)= R_0(1+\varepsilon_2 \cos 2 \varphi)$ in polar coordinates $(r, \varphi)$ with mean radius $R_0$ and deformation parameter $\varepsilon_2=0.05$ indicating a very slight deviation from the circular shape. b) Representation of the trajectories in the Poincaré surface of section (PSOS) reveals a clear picture characterized by the formation of chains of elliptical fixed points that structure the phase space. Colors are a guide to the eye and may differ in a) and b).}
\end{figure}

One could argue that we can distinguish very well between regular and chaotic trajectories in real space by the visual impression, at least in the classical case. However, this is a qualitative, individual judgment. Moreover, it leaves out important information that we actually do have at hand: the particle's momentum. It is encoded in the momentum component parallel to the interface tangent at the reflection point and conserved upon scattering. In (the isotropic) case it is nothing else than the sine of the angle of incidence $\chi$. An example illustrating how a phase space representation may enhance the understanding of the system's dynamics is shown in Fig.~\ref{fig_phsp_class}.

\begin{figure}
	\includegraphics[width=1.0\textwidth]{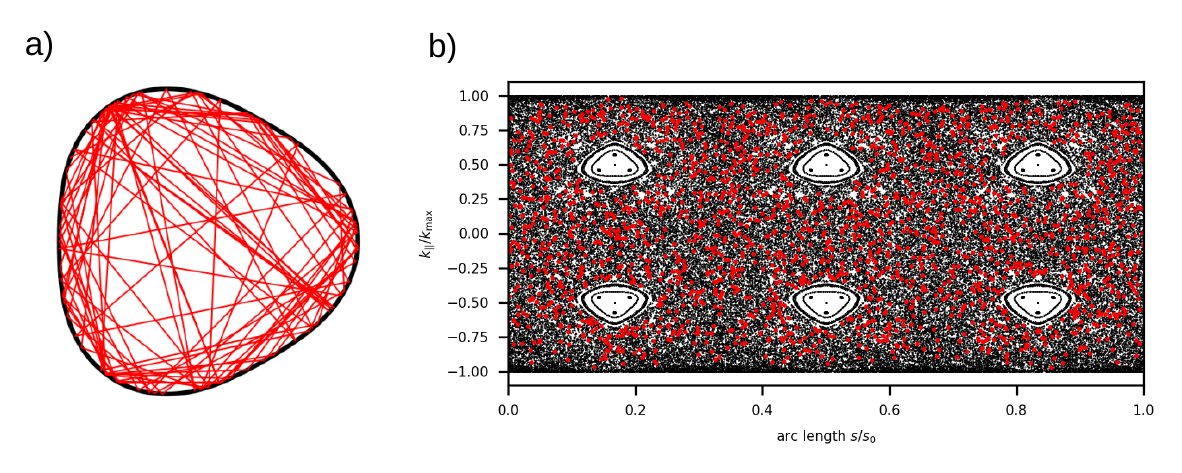}
	\label{fig_phsp_class-chaot_onigiri}
	\caption{a) A chaotic trajectory in an onigiri-shaped cavity, $r(\varphi)= R_0(1+\varepsilon_3 \cos 3 \varphi)$ in polar coordinates $(r, \varphi)$ with mean radius $R_0$ and deformation parameter $\varepsilon_3=0.08$ followed for several hundred of reflections will eventually fill the real space ergodic. b) The red dots indicate the first 1000 reflections of this trajectory in the PSOS 
	(black dots/lines). They fill the chaotic part of phase space ergodic, however, stable islands are not accessible to a chaotic trajectory. }
\end{figure}


\begin{figure}
	\includegraphics[width=1.0\textwidth]{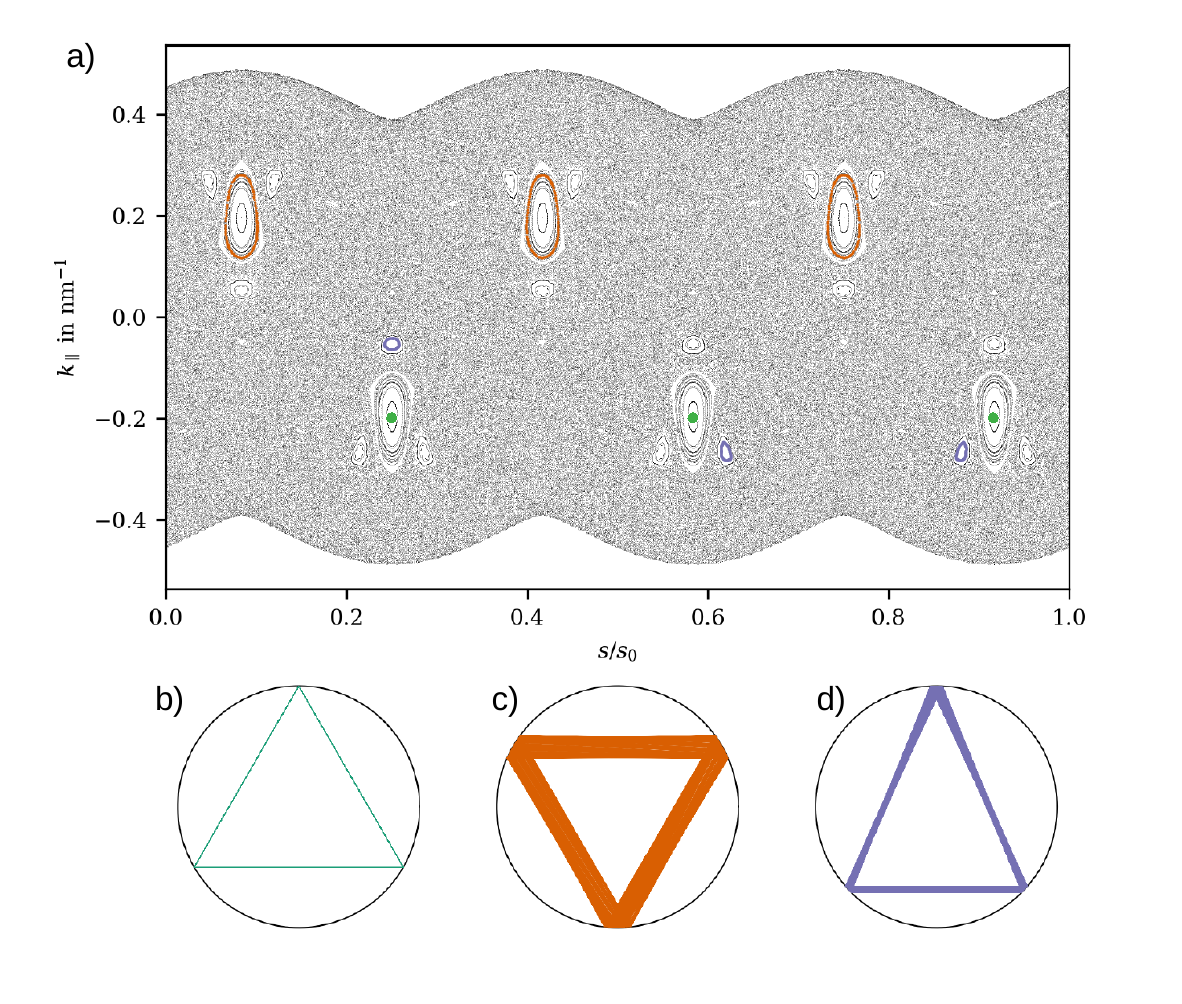}
	\label{fig_phsp_graph}
	\caption{a) 
		PSOS for an anisotropic BLG billiard with circular geometry. The chains of three islands, shifted between the clockwise and counterclockwise sector are characteristic and represent the three preferred propagation directions. The curly line of the maximum/minimum value of $k_\parallel$ as a function of the arc length $s/s_0$ reflect the different radii of a non-circular Fermi line in momentum space. b),c) and d) illustrate typical trajectories that are color coded in a).}  
\end{figure}

So here are two points in favour of extending the consideration to momentum space:

Let's start with the classical case. The characteristics of the integrable case is the conservation of angular momentum, the sine of the angle of incidence, $\sin \chi$, is conserved. A map where this is directly evident would be nice indeed. The more as chaotic trajectories can then be directly distinguished by the non-conservation of angular momentum. 

Such a map exists and is well known as the Poincaré surface of section (PSOS). The PSOS maps the four-dimensional phase space of a two-dimensional (billiard) system on two dimension by recording the information at the system's boundary only: Whenever a reflection event occurs, the position $s$ along the system's boundary is taken together with $\sin \chi$ as a measure for the momentum parallel the system's boundary (tangent).
Thus trajectories in integrable systems with constant $\sin \chi$ stand out as horizontal lines. 

Regular trajectories occur not only in integrable systems, but also in mixed systems. When an integrable system is perturbed, the Poincaré-Birkhoff theorem  \cite{lichtenberglieberman1992} takes over: Regular trajectories decay into a chain of elliptic and hyperbolic fixed points. Both represent periodic trajectories, but these are either stable or unstable. The stable periodic orbits survive perturbations away from the fixed point to a certain extent. This means that around the elliptic fixed point
stable islands are formed. They, too, are visible in the PSOS right away, see the example in Fig.~\ref{fig_phsp_class}. 

What about chaotic trajectories? There is no conserved quantity and no memory of the integrable system. Chaotic trajectories fill the space between regular islands 
ergodic, cf.~Fig.~\ref{fig_phsp_class-chaot_onigiri}. In fact, when we start a chaotic trajectory and wait for a sufficient long time, it will explore all of the chaotic part of the phase space of the system, see Fig.~\ref{fig_phsp_class-chaot_onigiri}b). In contrast, to get an impression of the regular part of the phase space, a sufficiently large number of trajectories has to be 
started individually. Notice that the chaotic trajectory in  Fig.~\ref{fig_phsp_class-chaot_onigiri} does, consequently, not reach the regular islands that remain white and reserved for stable trajectories ín phase space, while the tracing in real space fills out the whole cavity space.

We close the discussion of the classical case by a remark on a phase space most of you will know, for example from the analytical mechanics course: the one-dimensional pendulum as a Hamiltonian system (see \cite{pendulum} for an illustration). 
 Here, the phase space is two-dimensional in canonical coordinates spanned by the angle of deflection $\phi$ and by the velocity $\dot{\phi}$, resulting in the well-known representation of oscillatory motion in the form of (concentric) ellipses that become the larger the larger the deflection amplitude is. The message of the concentric, non-intersecting ellipses is that the velocity $\dot{\phi}$ is zero at the turning points, and takes its maximum absolute value for ${\phi}=0$ (with the different signs corresponding to the opposite velocity directions needed to form the oscillatory motion), and the maximum $|\dot{\phi}|$ increases with the deflection amplitude.
 When the (maximum) $|\dot{\phi}|$ or the (maximum) deflection amplitude ${\phi}$ become too large, the pendulum dynamics changes qualitatively from oscillatory motion to rotation. Its phase-space fingerprint are 
 (curly) horizontal lines of either positive or negative $\dot{\phi}$ that correspond to counterclockwise or clockwise rotation, respectively. 
 
 The pendulum's phase space is structured by fixed points: a stable (or elliptic) fixed point, the center of the ellipses, that corresponds to the resting (hanging) pendulum. Stable means that a small distortion of the pendulum leaves it close to the fixed point. Symmetrically to this stable fixed point there is an unstable (or hyperbolic) fixed point corresponding to the "standing" pendulum (assuming a stiff thread): the slightest perturbation will drive the system away from this fixed point and bring it into a rotational motion (as there is energy conservation, and in particular no friction in Hamiltonian systems).

 The concept we use here to represent the dynamics of two-dimensional billiards is very similar: we choose one spatial and one momentum coordinate, however, now out of a four-dimensional phase space that we reduce to two dimensions by sticking to the information at the system's boundary, i.e.~at the reflection points. Periodic orbits are represented by regular islands consisting of concentric ellipses around an elliptic (stable) fixed point. The number of islands signifies the number of reflections needed to return to (the vicinity) of the starting point. So three islands correspond to triangular orbits, two islands to bouncing ball orbits, and so on. The elliptic signature arises, in our time-discrete surface-of-section method  when the original reflection point is not precisely hit 
after one roundtrip (incommensurate initial conditions), while in the pendulum case discussed above the ellipse represents the time-continuous monitoring of the oscillator coordinates $\phi$ and $\dot{\phi}$. (There is no free lunch, and less so when going from a one- to an two-dimensional system. However, the time-discrete version allows for a convenient and very useful representation.)

Figure \ref{fig_phsp_graph} illustrates the 
PSOS for a circular bilayer graphene billiard, an anisotropic system. We observe chains of three islands as in the onigiri case discussed in Fig.~\ref{fig_phsp_class-chaot_onigiri}, but their origin is different: The anisotropy itself represents a perturbation to the isotropic circular (and therefore integrable) system, that results, very much in the spirit of the Poincarè-Birkhoff theorem \cite{lichtenberglieberman1992}, in a sequence of elliptic and hyperbolic fixed points. The resulting structure of the PSOS is exemplarily shown in Fig.~\ref{fig_phsp_graph}a) with typical trajectories shown in Fig.~\ref{fig_phsp_graph}b), c) and d). We restrict ourselves to just one remarkable feature of bilayer graphene physics here (and refer the reader to \cite{revgraphene,revgraphMcCann,BLG1,BLG2,BLG3,BLG4} for details such as the tunable external parameters to realize such a situation), namely the existence of three preferred propagation directions with 120$^o$ angles between them. These can be combined to form a triangular trajectory such a in Fig.~\ref{fig_phsp_graph}b). Its phase space signature are stable fixed points right in the center of the three lower islands, corresponding to a fixed sense of (clockwise) rotation. However, the three preferred directions can be put together in another triangular orbit, with opposite sense of rotation, cf.~Fig.~\ref{fig_phsp_graph}c) with signature in the upper half plane of the PSOS. Here, the fixed point is not perfectly hit, but more importantly, the trajectory is rotated in space and therefore the islands of the triangular orbits appear shift in the PSOS - in stark contrast to the isotropic momentum space with the onigiri-type deformation in real space in Fig.~\ref{fig_phsp_class-chaot_onigiri}.  However, the situation is more symmetric than it appears on first sight: the other triangular orbit does exist in the form  of unstable fixed points right in between the stable fixed points. From the theoretical point of view of nonlinear dynamics, this is expected from the Poincaré-Birkhoff theorem. From a more practical perspective, the wave vectors needed to form these cavities do exist, even though they not preferred directions.

The avalanche of the formation of chains of elliptical and hyperbolic fixed points suggested by the Poincaré-Birkhoff theorem holds for the anisotropic case as well, and Fig.~\ref{fig_phsp_graph}d) represents a trajectory that is hosted by (stable) islands formed around the central island corresponding to the symmetric triangular-orbit. 

Figure \ref{fig_phsp_graph}a) may suggest chiral properties of the anisotropic BLG billiard since the clockwise-counterclockwise (left-right) symmetry appears to be broken. It is crucial to note that we discussed the situation so far for only one of the two different so-called graphene valleys \cite{revgraphMcCann,revgraphene}. Then the chiral symmetry is broken indeed, but cured once the full system with both valleys are taken into account.

\section{Husimi functions for open billiard systems}

\begin{figure}
	\includegraphics[width=1.0\textwidth]{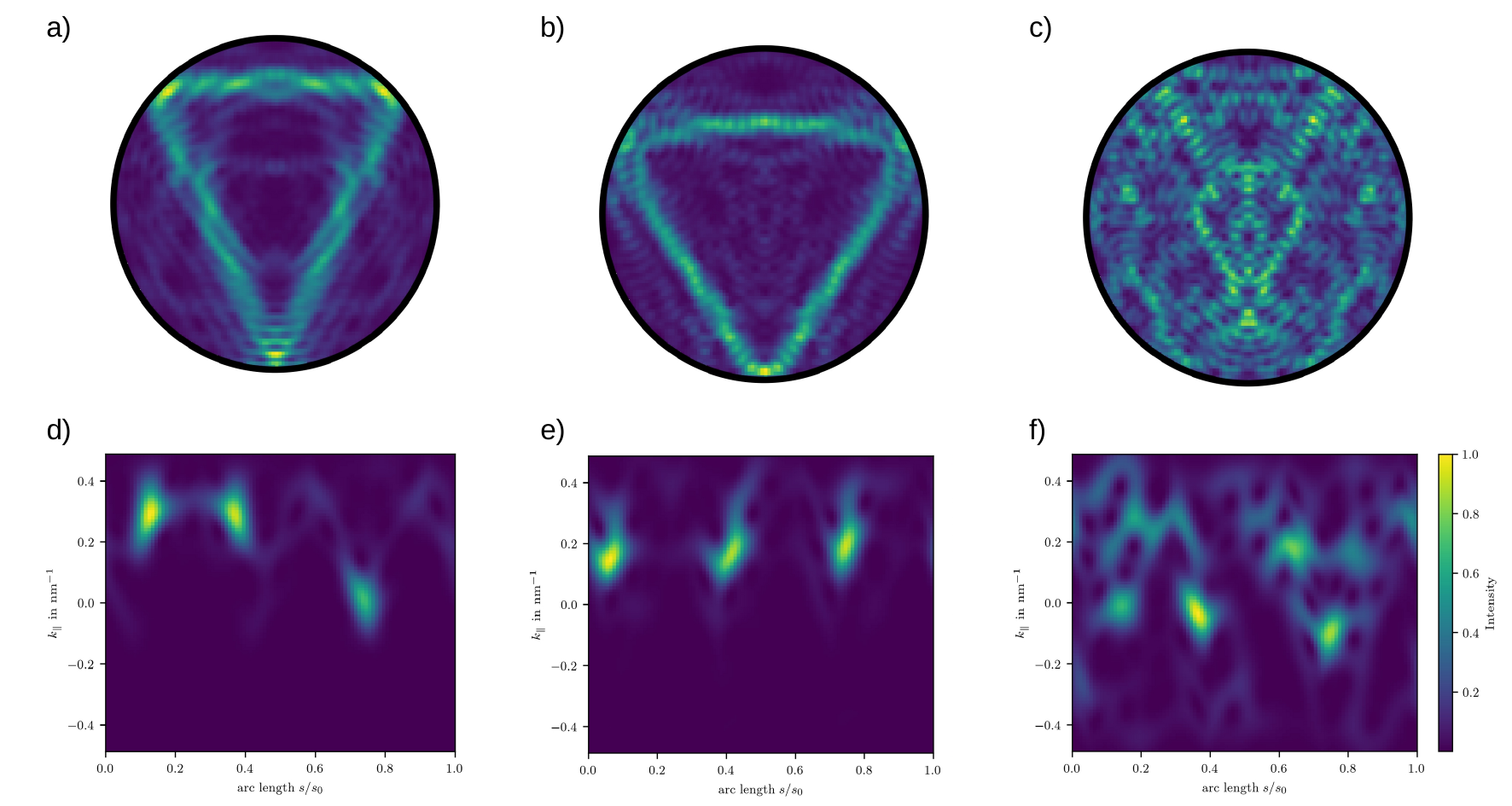}
	\label{fig_phsp_husimi}
	\caption{a),b), and c) show sample wave functions of a circular bilayer graphene billiard and d),e), and f) the corresponding Husimi functions $H^\mathrm{inc}_1(k_\parallel^1=k_\parallel)$ in color scale (far right). The difference in the phase space fingerprints is clearly visible, with a nice qualitative correspondence of the triangular-type modes in a), b) and d), e) to the trajectory tracing signatures in the PSOS shown in Fig.~\ref{fig_phsp_graph} (up to $K^{\pm}$-valley related symmetries). The panels c),f) illustrate a more chaotic resonance. The data shown is obtained for the so-called $K^+$ valley.}
\end{figure}

In order to take full advantage of 
this useful concept of the phase-space representation we need a way to transfer it into the quantum mechanical situation. 
The above-mentioned concept of the Poincaré surface of section for a billiard systems is based on knowledge of precise reflection points that do not exist any more. We shall see that the means of choice to overcome this problem is the Husimi function that we introduce in the following section. It can be thought of as measuring the wave function by taking the overlap with a coherent wave packet centered at a certain position and representing a specific (angular) momentum. The information in the PSOS 
will be smeared out as well, but this is to be expected. Doing the mapping, or "measurement" of the wave function with a coherent state, i.e. a minimum uncertainty wave packet, ensures that best possible resolution under quantum mechanical conditions is achieved.

The Husimi function, first introduced by K\^{o}di Husimi \cite{husimi_kodiHusimi}, as well as the Wigner function (that can take negative values  and shall not be further considered here\cite{phasespacequantummechanics}), have proven to be established tools in the analysis of quantum dynamics, quantum-classical correspondence as well as in the field of quantum chaos \cite{baecker1_hus,baecker2,baecker3,koreanerStadium,review_phsp_Curtright_2012}. The Husimi function for closed (hard wall) billiard systems decomposes the wave function's (normal) derivative at a point $s_0$ along the 
system boundary into its angular momentum component by computing the overlap with a coherent state $\xi (s,m)$ with spatial coordinate $s$ along the boundary (normalized by the circumference $s_0$ of the cavity) and angular momentum $m=m(\sin \chi)$ that is centered around $s$. Notice that the wave function itself drops to zero at the system boundary. Using the relation $\sin \chi = m /(nkR_0)$ between angular momentum and angle of incidence, and introducing $k_\parallel^j = k_0 \sin \chi^j$, the momentum component parallel to the interface tangent at the reflection point at interface side $j$,  the coherent state can be written as
\begin{equation}
	\xi(S_p;s,k_\parallel^j) = (\sigma \pi)^{-\frac{1}{4}} \sum_l e^{-\frac{1}{2\sigma}(S_p-s + l )^2 - i k_\parallel^j (S_p - s +  l ) } \:.
\end{equation}
It is periodic in $S_p$, and $\sigma$ fixes the uncertainty in $s$ (and thus in $k_\parallel$) direction.

In an open system, both the wave function and its derivative are non-vanishing at the system boundary. At the same time, part of the wave function amplitude might leave the system in a reflection event. In total, there are four trajectory parts contributing, namely incoming and outgoing both inside ($j=0$) and outside ($j=1$) of the boundary. All should have their analogues in a Husimi representation. It turns out that both issues can be solved simultaneously by combining the Husimi function based on the wave functions and its derivative (with a phase factor of $i$), respectively, to a generalized Husimi function for open systems \cite{husimiepl}. 

The formulae for these four Husimi functions, incident (inc) and emerging (em), were derived in \cite{husimiepl} and read
\begin{equation}
	H^\mathrm{inc(em)}_j(s, k_\parallel^j) = \frac{k}{2 \pi} \left | - {\cal F}_j \; h(s, k_\parallel^j) + (-) \frac{i}{k {\cal F}_j} \; h'(s, k_\parallel^j)\right |^2 
\end{equation}
with the weighting factor ${\cal F}_j= \sqrt{n \cos \chi_j}$ where $n$ is the refractive index of the optically thicker medium assumed to be embedded in vacuum or air. In the more general case of, for example, anisotropic cavities, $n$ can be effectively defined (approximated) as the ratio (of the maximum value, taken from the center of the Fermi line) of the wave vectors in the inner and outer media. The individual Husimi functions $h$ and $h'$ are based on the wave function and its (normal) derivative, respectively, as
\begin{eqnarray}
	h(s, k_\parallel^j) = \oint \psi(s) \; \xi(S_p,s,k_\parallel^j) dS_p \\
	h'(s, k_\parallel^j) = \oint \psi'(s)\; \xi(S_p,s,k_\parallel^j) dS_p	
\end{eqnarray}

Husimi functions are illustrated in \cite{annbill} and Fig.~\ref{fig_raywave}c) for optical cavities and in Fig.~\ref{fig_phsp_husimi} for an circular bilayer graphene (BLG) billiard \cite{BLG1,BLG2,BLG3, BLG4}, focusing on the typically considered inside-incident Husimi function $H^\mathrm{inc}_1$. Bilayer graphene exhibits the so-called trigonal warping in the dispersion relation that yields a preference for triangular orbits even in a circular cavity of this anisotropic material. The two different triangular wave functions shown in Fig.~\ref{fig_phsp_husimi}a),b) result in rather different Husimi signatures, cf.~Fig.~\ref{fig_phsp_husimi}d),e). We point out the dominance of positive $k_\parallel$ and the clearly visible difference in the angles of incidence $\chi$, where $\sin \chi$  is replaced by the conserved momentum $k_\parallel / k_0$, for the two triangular-type resonances shown in Fig.~\ref{fig_phsp_husimi}a) and b), respectively. The chaotic wave function shown in Fig.~\ref{fig_phsp_husimi}c) with its Husimi function in Fig.~\ref{fig_phsp_husimi}f) possesses, in contrast, a much broader intensity distribution both in real and momentum space.








\section{Lessons learned from the phase-space perspective}

In this chapter we will illustrate the capability of a phase-space approach to quantum chaos in various examples, ranging 
from nonlinear dynamic aspects via semiclassical effects to lasing cavities. It cannot be a full-depth discussion, and we will focus on some of the original literature and some review articles, without being able or even attempting to be complete. Rather, we would like to give some overview and hopefully a starting point into this fascinating world. 

\begin{figure}
	\includegraphics[width=1.0\textwidth]{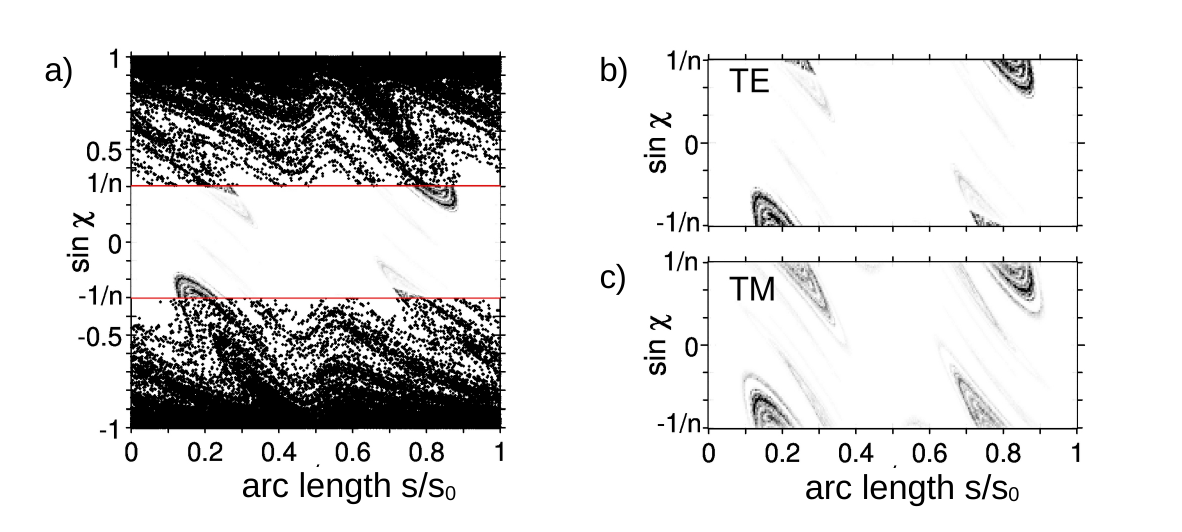}
	\label{fig_instabmanif}
	\caption{From the unstable manifold of closed billiard systems to the steady probability distribution of open photonic resonators, here a lima\c{c}on cavity,  $r(\varphi)= R_0(1+\varepsilon_1 \cos \varphi)$, with geometry parameter $\varepsilon_1=0.43$ and refractive index $n=3.3$. a) Pre-images of points in the PSOS that had crossed the critical lines $\sin \chi_c = \pm 1/n$ (indicated in red). Within the leaky region, that is, inside the red lines, the intensity is weighted by the Fresnel reflection coefficients for TE polarised light. In b), the leaky region is enlarged. In c), same as for b) but for TM polarised light. }
\end{figure}

\begin{figure}
	\includegraphics[width=1.0\textwidth]{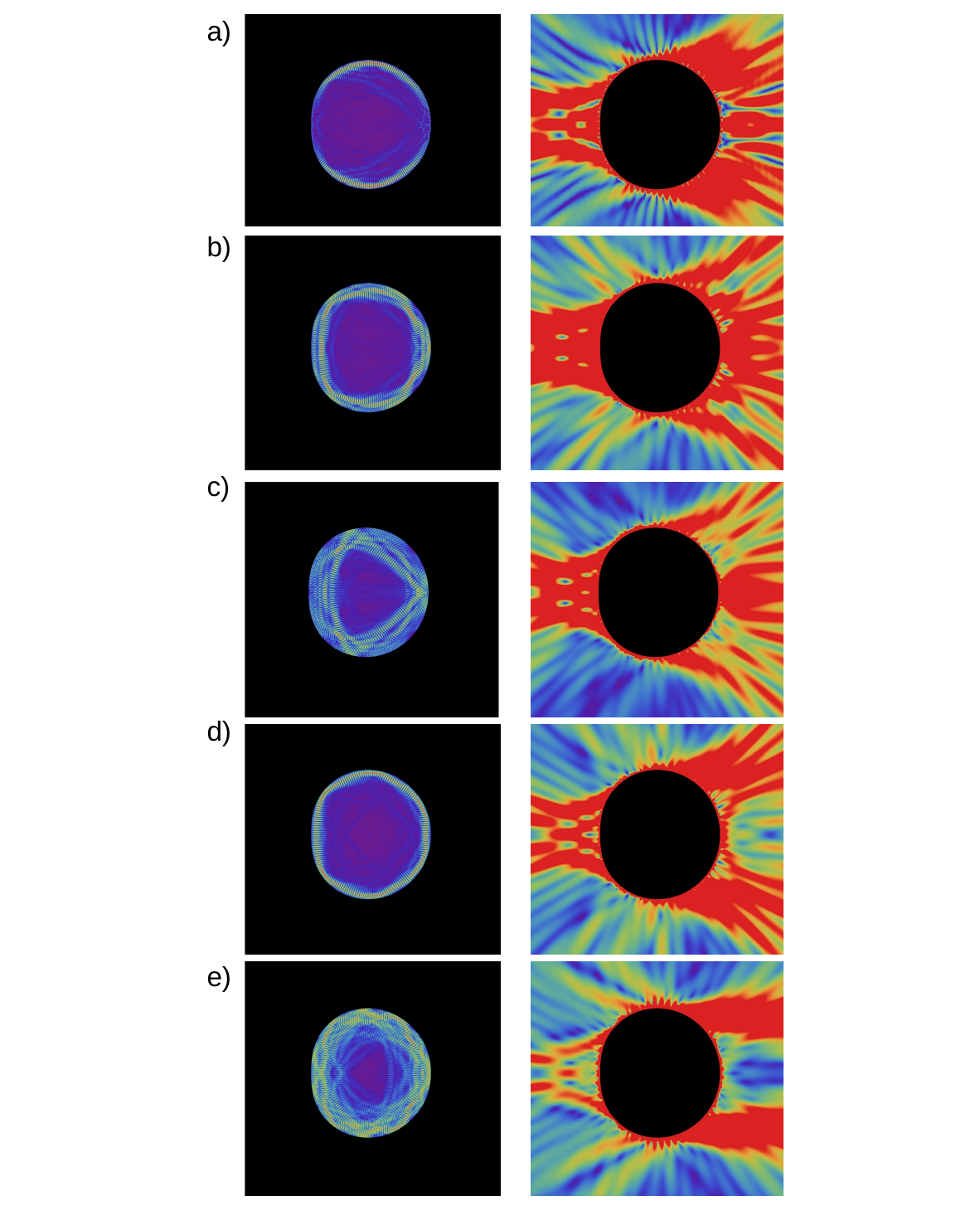}
	\label{fig_examTE}
	\caption{Qualitative illustration of the unstable-manifold induced robustness of the emission properties of mesoscopic optical cavities, here for a lima\c{c}on resonator with ${\varepsilon}_1=0.43$ and refractive index $n=3.3$ for TE polarised resonances. Left panels of a),b),c),d), and e) show intensity of resonances inside the cavity, using comparable scaled vacuum wavenumbers Re[$k R_0$] $\approx$ 20 and different $Q$ factors,
		a) $Q=$ 299296, b) $Q=$ 228698, c) $Q=$ 80966, d) $Q=$ 27397, and e) $Q=$ 5236. The right panels show the outside intensity with red indicating high values.  
	}
\end{figure}

\subsection{Quantum resolution of classical structures in phase space}

There is no resolution limit for the phase space in classical mechanics, and both position and momentum can, in principle, be given with arbitrary precision. In contrast, Heisenberg's uncertainty principle changes the situation in quantum and wave mechanics: The product of the resolution in position, $\delta x$, and momentum, $\Delta p$, is of the order of Planck's constant $\hbar$, or, in the case of (electromagnetic) waves, of the order of the wavelength or inverse wave number $k$. Notice that such a relation is familiar in optics for example from Gaussian beams, where an uncertainty-type relation holds between the waist of the beam and its divergence that is inversely proportional to its beam waist \cite{photonikbuch}. 

As a consequence, structures of the classical phase space can only resolved to a certain, wavelength dependent extent by resonances \cite{BOHIGASTOMSOVICHULLMO}. Small islands as well as other structures might be `not visible` for resonances with too large wavelengths \cite{shim2}. In turn, if the distance between adjacent islands remains smaller than the uncertainty limit, they appear as a continuous chain to resonances of the corresponding (and larger) wavelength \cite{shim1,partialbarriers,BLG2}.

There is one topic that is not considered here in more detail because it would require at least one chapter in its own just to state the literature: Scarring. There are resonances living along trajectories that are classically unstable! In \cite{heller_scars}, the author Rick Heller coined the term scars of periodic orbits as one would not expect them to exist such resonances in the first place. So it turns out that things are not ideal, or much better to say, more interesting than expected. Scars became and reamin a popular and intense field of research, including the field of optical microcavities where also a more general localisation of modes is under discussion. 


\subsection{Dynamical and resonance assisted tunneling}

Besides the uncertainty relation, there is another phenomenon characteristic for quantum behaviour: tunneling. Conventionally, it refers to a quantum particle penetrating a barrier despite the fact that energy conservation would not allow it so that the classical pendant is forbidden. Here, we observe yet another variant, the chaos-assisted dynamical tunneling \cite{chassistunn}. It refers to the situation where there exist two classically strictly separated movements that are ultimately connected quantum (wave) mechanically, and this connection is enhanced in the presence of chaos. One example are clockwise and counter-clockwise propgating whispering gallery modes in a circular cavity that are clearly separated and possess exactly the same energy. If now the cavity is slightly deformed and thereby chaos is induced, their energies will split a little bit. 
The mechanism behind is their coupling by dynamical tunneling. In the presence of (some) chaos, 
tunneling does not have to take place from, say, one chain of islands to another one, but rather from the chain of islands to the surrounding chaotic layer that ensures fast transport across phase space to another (chain of) islands that is again reached by (conventionally) tunneling. Sometimes such a  situation is referred to as "regular islands connected by the chaotic sea". It underlines that this mechanisms requires a mixed system (or a system far from integrability \cite{chassistunn}) to be effective. Concerning more generic and diverse 
aspects of transport in the presence of chaotic dynamics, we would like to refer the reader to the chapter Chaotic Dynamics and Quantum Transport by
Andrey R. Kolovsky in this volume. 

\subsection{Microdisk laser as applications}
\label{sec_microlaser}

There was intense research in the field of quantum chaos in the 1990s and early 2000s. One direction of investigations concerned optical microcavities \cite{VahalaBook,xiaoAsymmetricResonantCavities2010} and the emergence of mesoscopic optics in general, especially the wish (for me, a young researcher back then, it felt a bit like the holy grail) to have a microdisk laser with directional emission available. 

Microdisk cavities formed in semiconductor heterostructures could be build in essentially all desired shapes then and possessed very high $Q$ factors resulting from the high quality of the side walls - their smoothness ensured scatter-free, ballistic propagation and confinement by total internal reflection. Classically, total internal reflection occurs when the angle of incidence $\chi$ is large than the critical angle $\chi_c$ given as $\sin \chi_c = 1/n$. In the quantum or wave situation, there is tunneling or evanescent escape, described by the imaginary part of the resonance wave number that in turn can be related to the $Q$ factor or a generalized Fresnel reflection coefficient (see Sec.~\ref{sec_semiclass} below).  

But a circular disk microlaser would send out (tunneling) light in an isotropic manner, not at all like a laser. Before long, there had been many innovative and successful ideas presented to overcome this issue, cf.~the reviews \cite{revunidir1,revunidir2} for further reading. One option is using a special geometric shape, the so-called lima\c{c}on cavity,  $r(\varphi)= R_0(1+\varepsilon_1 \cos  \varphi)$ in polar coordinates with geometry parameter $\varepsilon_1$,
 first theoretically predicted \cite{limacon}, and experimentally confirmed by several groups worldwide \cite{limacon_Cao2009,limacon_Kim2009,limacon_Susumu_Taka2009,limacon_NJPhys,limacon_Albert2012}, with a remarkable agreement between ray and wave simulation as well as with experiments, although all three results were obtained in very different wavelength regimes. 
 
The origin of this robust and resonance independent directional emission lies actually in the unstable manifold of the hardwall billiard counterpart, cf.~Fig.~\ref{fig_instabmanif}. A chaotic Hamiltonian system such as the lima{\c{c}}on has to have an unstable direction in phase space (otherwise no chaos), and a stable (contracting) direction to fulfill Liouville's theorem. The result are the filament-like structures in phase space (especially in the PSOS) typical for chaotic systems. They are visible in Fig.~\ref{fig_instabmanif}a) as black dots that were obtained as the pre-images of 
 trajectory reflection points after they had first violated the condition for total internal reflection.  

In open billiard systems, the unstable manifold is replaced by the so-called steady probability distribution \cite{unstabmanif}) and  
has to be weighted by the Fresnel reflection coefficients within the critical lines $\sin \chi_c = \pm 1/n$ as is shown in Fig.~\ref{fig_instabmanif}b),c) for the two possible polarisations (TE and TM polarisation, respectively, with the magnetic and the electric field, respectively, pointing in $z$-direction perpendicular to the cavity plane). 
Therefore, the value of the refractive index plays an important role as it defines the position of the critical line for total internal reflection and thus which part of the unstable manifold is not confined by total internal reflection any more. It turns out that for the lima\c{c}on cavity with a refractive index $n$ around 3.3 and a geometry parameter $\varepsilon_1 \approx 0.43$, and in a certain range around these parameters, the few filaments reaching into the leaky region yield indeed directional emission. For smaller refractive indices such a glass, another geometry has to be used, the so-called shortegg \cite{shortegg} or onigiri shape. In both cases, the properties of the steady probability distribution combine to yield a far-field emission that is highly non-uniform and prefers very few directions. The Brewster-angle feature in the Fresnel transmission coefficient for TE polarised light yields typically the better directionality in the far field because it functional dependence on the angle of incidence mimics a step-like (all or nothing) function, see the difference in Fig.~\ref{fig_instabmanif}b) and c). 

The importance of the steady-probability distribution for the far-field emission implies a robust, universal and resonance independent behaviour. This is illustrated in Fig.~\ref{fig_examTE} where a number of resonances are shown with their resonance intensity morphology (left panels) and the corresponding near-field emission. Whereas the wave patterns vary and are ordered according to decreasing $Q$ factors that can be derived from the imaginary part of the scaled wavenumber $k R_0$ as 
 \begin{equation}
 		Q = -\frac{\mathrm{Re}(k R_0)}{2 \mathrm{Im}(k R_0)},
 	\end{equation}
the unstable-manifold inspired emission patterns shown in the right panels are rather similar. Note that they do, however, show the characteristic mesoscopic fluctuations resulting from (tiny) interference -originated differences in nearby resonances that are to be expected. This important property of mesoscopic systems is studied in detail in Ref.~\cite{limacon_Susumu_Taka2009} where it is shown that averaging over many resonances improves the agreement with both the ray model and experimental results because resonance-specific features are taken out.  

For the investigation of active, lasing cavities with amplification described for example with the Schrödinger-Bloch\cite{SBmodelPRL,FukushimaT_JOS_2004} model, the existence of four Husimi functions is particularly useful. Considering the incoming and outgoing Husimi functions at a certain (reflection) point along the cavity boundary, and following it to the next reflection event, allows one to analyse and understand the amplification process within the cavity, and consequently the laser emission \cite{dynamichusimi}. This holds also in the case of space-selective pumping. 


\subsection{Ray-wave correspondence, extended ray picture, and semiclassical effects}
\label{sec_semiclass}


\begin{figure}
	\includegraphics[width=1.0\textwidth]{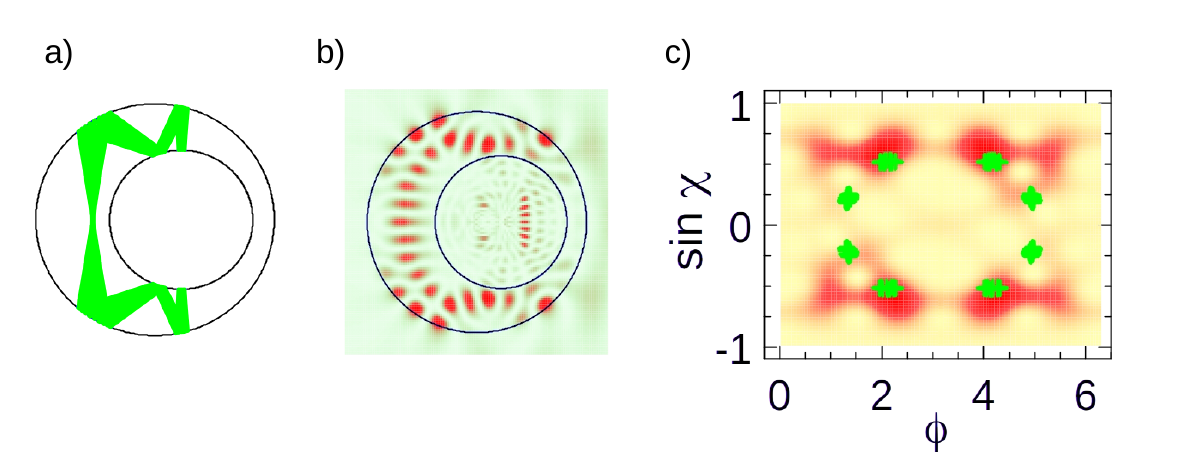}
	\label{fig_raywave}
	\caption{Ray-wave correspondence in the optical annular billiard. a) Example of a stable ray trajectory in the hard wall system. b) Wave solution obtained from a scattering matrix approach \cite{annbill} for a system with refractive index $n_1=3$ in the annular region and $n_2=6$ in the inner region, embedded in air. While a) and b) show the real space, c) compares rays and waves in momentum space by superposing the PSOS signature of a) (green crosses) and the Husimi signature (inside incoming) of b) (red colors denoting high intensity) highlighting similarities and differences. }
\end{figure}

\begin{figure}
	\includegraphics[width=.90\textwidth]{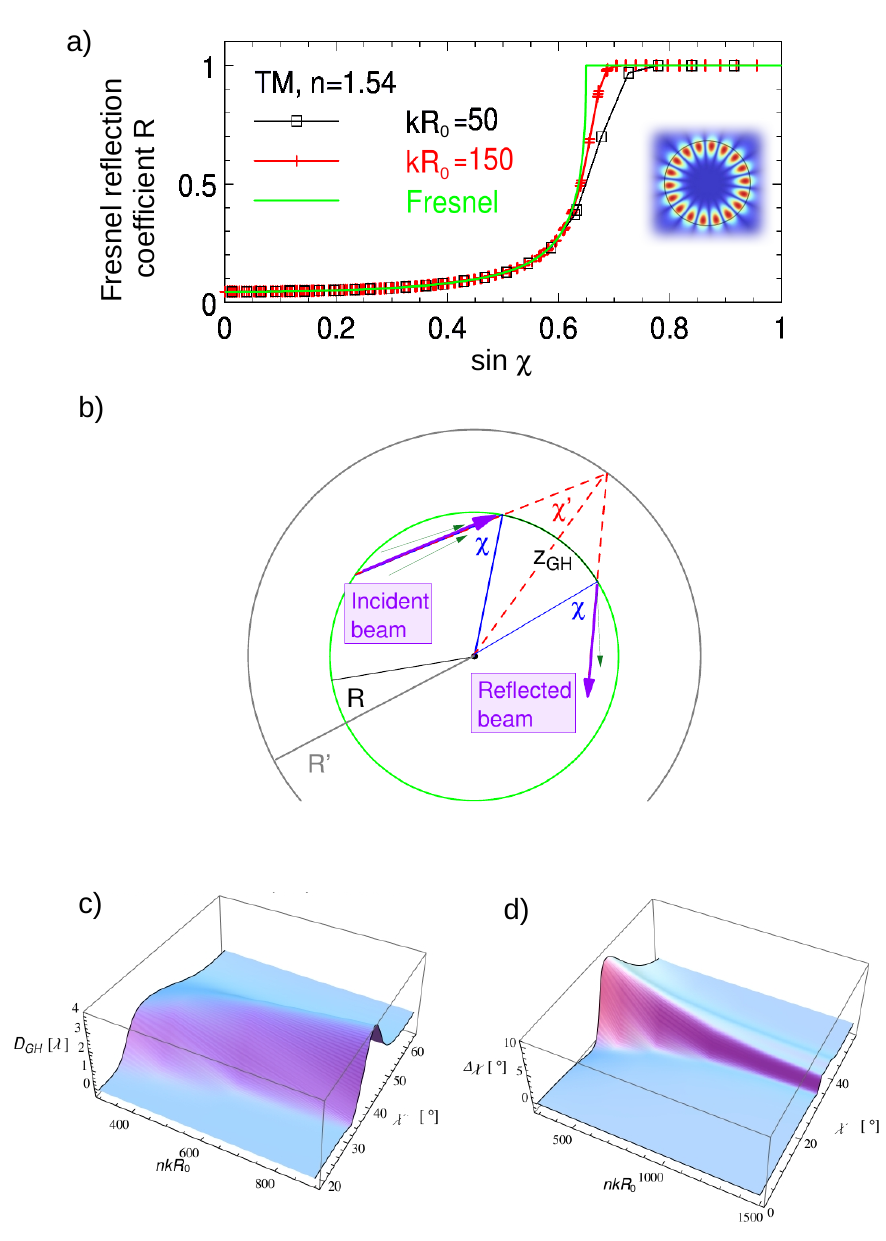}
	\label{fig_semiclass}
	\caption{Semiclassical corrections to the ray picture. a) Fresnel reflection coefficient for Tm polarised light and $n=1.54$ for ray optics (green) and two different wave numbers corresponding to two different wavelengths. The larger the wavelength $2\pi/k$, the larger are the deviations from the ray picture expectation. b) Goos-Hänchen shift, a lateral shift $z_{GH}$ along the reflecting interface upon total internal reflection, partly explains this behaviour, see text for details. c) Scaled Goos-Hänchen $D_{GH}$ shift as function of curvature and angle of incidence. d) same as c) but for Fresnel filtering $\Delta \chi$.  }
\end{figure}

Ray-wave correspondence has been established as a straightforward and versatile semiclassical tool in the investigation of ballistic mesoscopic billiard systems. It assumes that the light ray or particle moves along straight lines, i.e. ballistically and without scattering (that would dominate in disordered systems) and such that the phase information is kept by its wave analogue. Note that the underlying (atomic scale) microstructure of the material is not considered as it is not resolved: the wavelength of the particle or ray has to be larger than the lattice constant but smaller than the system size. 

 An illustration of ray-wave correspondence is shown in Fig.~\ref{fig_raywave} where a ray and wave solution are compared in real and momentum space, and we refer the reader also to examples discussed above. 
 The success of ray-wave, or more generally, particle-wave correspondence is remarkable and extends beyond mesoscopic optic to electronic systems such as Dirac fermion optics in graphene \cite{DiracFermionOptics} or to anisotropic systems discussed in below in Sec.~\ref{sec_aniso}.

However, deviations from the ray (tracing) picture are expected because of the inevitable presence of interference effects on the wave side, and do exist in various forms. The differences between the ray trajectory and the resonances, characterized by an sequence of intensity maxima, are a direct and prominent example. The slight differences between overall similar resonance patterns are also induced by slight and resonance-dependent variations in the interference "affairs".

Another example is illustrated in Fig.~\ref{fig_semiclass}a) and shows deviations between the Fresnel reflection coefficient (for TM polarised light) well-known in ray optics for the reflection at an planar interface boundary and the generalized Fresnel coefficient deduced from the imaginary part of the resonance wave number,
\begin{equation}
	R=\exp(-4n \mathrm{Im}(k R_0) \cos \chi)\:.
\end{equation}    
Where as the onset of total internal reflection ($R=1$) is sharp in the ray optics situation, the curves appear smoothed for the resonances (one example wave pattern is shown in the inset), and the more the more larger the wavelength $\lambda= 2 \pi/k$ or the smaller the wavenumber $k R_0$ is. 

This can be understood qualitatively and semiquantitatively \cite{Stockschläder_2014} by means of the Goos-Hänchen shift \cite{GHS,fresnel1}. To this end we consider the evanescent wave that is present in the case of total internal reflection, implying that the light beam penetrates of the order of the wavelength $\lambda$ into the optically thinner medium. In other words, the reflection (in the sense of the ray model) takes place at an effective interface of the order $\lambda$ "behind" the original boundary, cf.~\ref{fig_semiclass}b). So there is a lateral shift between incoming and reflected beam along the original interface, the so-called Goos-Hänchen shift \cite{GHS}.

At the curved interface of optical microcavities, this implies that the angles of incidence at the real, $\chi$ and effective,$\chi`$, interface are not the same (for concave interfaces, the angle of incidence is smaller at the effective interface, $\chi'< \chi$). This can be related to corrections to Fresnel's laws at curved interfaces \cite{fresnel1, fresnel2} for which adopted formulae could be derived \cite{fresnel1,fresnel3}. 

There is another semiclassical correction besides the Goos-Hänchen shift, referred to as Fresnel filtering \cite{ffstone}. Imagine a light beam incident at the critical angle on average. Now "beam", for example visualized as a typical Gaussian beam, means that there must be a certain distribution of angles of incidence when we assume a waist of the order of the wavelength (which is the only scale at hand). This is illustrated by the two small side rays in Fig.~\ref{fig_semiclass}b). Then "rays" (that constitute the beam) with angles of incidence smaller than the critical angle will still be refracted, while those with larger angles of incidence will be reflected. Consequently, the average angle of reflected rays is larger than the original angle of incidence, known as Fresnel filtering.  

It turns out that the phase-space perspective is particular useful in handling these two effects \cite{fresnel2} as it becomes evident that both effects act in different directions in phase space. Using incoming and outgoing Husimi function allows one to directly read-off their sizes \cite{fresnel2} and study their dependence on the boundary curvature \cite{Stockschläder_2014} or investigate them in an extended (classical) ray model \cite{fresnel-Altmann_2008}. The results for the curvature dependence are summarised in Fig.~\ref{fig_semiclass}c) for the Goos-Hänchen shift $D_{{GH}}$ and in Fig.~\ref{fig_semiclass} d) for the Fresnel filtering effect $\Delta \chi$ for a Gaussian beam of fixed waist width, i.e. a fixed spread of angles of incidence \cite{Stockschläder_2014}, TE polarisation, and $n$=1.5 the refractive index. While both the Goos-Hänchen and Fresnel filtering effect are largest around the critical angle, their dependence on the boundary curvature is opposite: The Goos-Hänchen shift decreases with increasing curvature (smaller $R_0$), while Fresnel filtering increases with any kind of curvature. 

We close this paragraph by noting the consequences of a ray picture that is extended by Goos-Hänchen shift and Fresnel filtering \cite{fresnel-Altmann_2008}. The dynamics in such an amended geometric optics turns out to be non-Hamiltonian in the sense that attractors and reppelors occur. This is caused by the Fresnel filtering rather than the Goos-Hähnchen shift that indeed can be thought of as reflection at an effective interface that does not change the Hamiltonian character of the dynamics.

\subsection{Anisotropic cavities}
\label{sec_aniso}

\begin{figure}
	\includegraphics[width=1.0\textwidth]{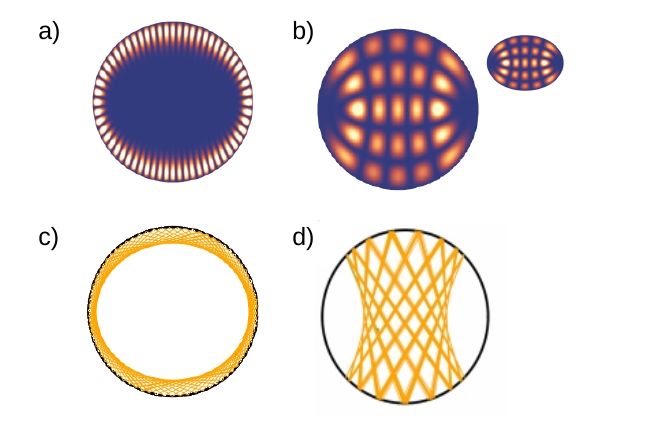}
	\label{fig_raywave_aniso}
	\caption{Ray-wave correspondence illustrated for the birefringent dielectric disk cavity with hard walls. While whispering-gallery-type resonances, see a) and c), remain present with a slight deformation, bouncing-ball-type orbits, see b) and d), emerge as new type of motion induced by the anisotropy. They follow the direction marked by the anisotropy (i.e. the extraordinary refractive index). The inset in panel b) shows a resonance mode of the isotropic ellipse that is related to the anisotropic disk by transformation optics. }
\end{figure}

Anisotropic cavities were already discussed above in the context of the bilayer graphene system. Here we add another system, namely the birefingent dielectric disk with hard walls \cite{anisodisk,Choi-Park:21}. As in the graphene case, the integrability of the system that is naively expected from its rotational invariant geometry is affected by the presence of anisotropies in phase space. The induced changes in the dynamics of the system are illustrated in Fig.~\ref{fig_raywave_aniso} by comparing corresponding wave patterns (panels a,b) in the upper row) and the ray tracing signatures (panels c,d) in the lower row), see also \cite{anisodisk}. We consider TE polarised light resonances, and the extraordinary (larger) refractive index  lies in the resonator plane, pointing into the $y$ axis. Note that other combinations of polarisation and orientation of the optical axis turn out to be less interesting, see \cite{anisodisk} for details.

While the whispering-gallery type modes known from the isotropic disk cavity are still present (though slightly deformed), cf.~Fig.~\ref{fig_raywave_aniso}a) and c), there is a new type of resonances emerging, bouncing-ball type patterns, cf.~Fig.~\ref{fig_raywave_aniso}b) and d). Those patterns are  known from the elliptic billiards \cite{WAALKENS199750}, and it can be shown (using for example transformation optics) that the anisotropic disk cavity behaves like an isotropic elliptical cavity \cite{anisodisk} (see \cite{Choi-Park:21} for the reciprocal case that maps an isotropic disk to an anisotropic elliptical cavity). The inset in Fig.~\ref{fig_raywave_aniso}b) shows the corresponding solution for the elliptic counterpart that has then to be stretched in $y$ direction. Ray-wave correspondence is confirmed.

\subsection{Wave resonances towards the ray limit}
Concerning ray-wave correspondence in optical cavity systems, an interesting question is how the transition to classical ray optics takes place when the wavelength is further and further reduced towards the ray limit of zero wavelength. To this end we refer the interested reader to recent work of the Ketzmerick group at TU Dresden, see \cite{ketz1,ketz2}, where the intimate (yet intricate) interplay between dynamical properties of the classical system and its quantum (including mesoscopic fluctuation) characteristics is discussed in detail. We point out that for wavelengths several orders of magnitude smaller than the system size, wave and ray-tracing solutions are hard to be distinguished qualitatively by eye (a special realisation of perfect ray-wave correspondence), while the quantitative analysis \cite{ketz1,ketz2} provides  a much deeper understanding of the interweaving of the classical and quantum aspects of our world.

\subsection{Transport, Sources, Leaking}
A brief extension of the discussion beyond the model system class "mesoscopic billiards for electrons and photons" to some more application oriented questions is in order. In experiments, there might be leads attached to the cavity such that a leaking system \cite{altmannleaking} results. In a similar manner, leads can act as sources of incoming particles, or there might be pointlike or extended sources placed within the cavity. All these situations can be conveniently described using the ray-tracing approach with appropriate initial conditions (for sources and incoupling leads) or dropout conditions (for outcoupling leads and leaks), and correspondence to the wave behaviour can be demonstrated 
\cite{entropysources}. 



\subsection{Non-Hermitian effects}

Finally, let us approach one aspect that brings us beyond the naive (yet useful and numerically cheap) ray and trajectory tracing picture. The focus of this chapter are open systems. These are, by nature, non-Hermitian systems characterized by complex eigenvalues. This enables access to the tremendously rich world of exceptional point and chirality physics \cite{revNonHermit,Bender_2007}. Exceptional points mark the coalescence of two resonance states in their real and imaginary parts upon variation of two external parameters such as refractive index and a geometry parameter (such as the position of nanoparticles attached to a cavity or the distance between two coupled cavities). In fact, exceptional points are rather common and useful, for example in sensors \cite{sensing_revWiersig}.

 
 Even if the exceptional point is not directly hit, its presence is visible via avoided resonance crossings in its vicinity. Exceptional points facilitate sensing applications due to the enhanced sensitivity in their vicinity \cite{sensing_revWiersig}. Their presence is intrinsically related to chirality, i.e., the predominance of a certain sense of rotation while the other angular momentum components are missing. Here the Husimi presentation turns out to be particularly useful as it allows one to directly access the angular momentum content of a given resonance: chirality means dominance of the intensity in the upper or lower half plane, see for example \cite{aplwithsile,epbosch}. At the exceptional point, the Husimi functions of the two coalescing resonance become indistinguishable. In a system of coupled deformed optical microcavities a whole chain of exceptional points (that occur at almost regular intercavity distances of several wavelengths) can form \cite{coupledlimacon2025} in the parameter space of refractive index and cavity distance. So exceptional points and constitute a network or backbone for the complex dynamics in open systems and enable new ways to applications.


\section{Summary}

Taking a phase-space perspective on quantum-classical correspondence can complement our understanding of quantum chaotic systems in many regards, in particular when (open or closed) billiard systems are concerned. In contrast to real space that focuses on the geometry of the systems and its characteristic trajectories, the phase space brings in the momentum information. In the case of mesoscopic billiards considered here, this is usually the angle of incidence at a reflection point, or the angular momentum, at the system boundary when the Poincaré surface of section is taken as the basis. Then a certain region in phase space corresponds to a certain position along the cavity boundary \emph{and} to a certain range of angles of incidence such that for example different senses of rotation can be directly distinguished.   
Understanding the systems's dynamics in phase space provides a valuable tool in designing systems with desired properties. 

Ray-wave correspondence proves to be a versatile and useful method in analysing the dynamics of quantum chaotic model systems such as billiards in real and phase space. Trajectory orbits and resonance wave patterns in real space are complemented by the Poincaré surface of section and the Husimi representation on the wave side.  The many insights that emerge from phase-space considerations of both electronic and photonic devices complement our picture and yield a more comprehensive insight into the fascinating physics of classical, quantum and mesoscopic systems.

\begin{ack}[Acknowledgments]\
We thank Lukas Seemann, Tom Rodemund, Silvan Stopp, Anna Skopnik, Samuel Schlötzer, Jana Lukin, Max Häßler, Malcolm Gebhardt, and Adrian Plenge for discussions and the results achieved during their thesis works at TU Chemnitz and Pia Stockschläder, Jakob Kreismann, and Sebastian Luhn at TU Ilmenau. We thank in particular Lukas Seemann for input to some of the figures.	
\end{ack}

\seealso{
Quantum analogues of exponential sensitivity: from Loschmidt echo to
Krylov complexity, 
Relativistic Quantum Chaos in Neutrino Billiards, 
The Quantum Kicked Rotor: A Paradigm of Quantum Chaos. Foundational aspects and new perspectives,
The role of classical periodic orbits in quantum many-body systems.}

\bibliographystyle{JHEP}%
\bibliography{lit_quch_phsp.bib}

\end{document}